\documentclass{svproc}

\usepackage{url}

\usepackage{amsmath, amssymb, graphicx}
\usepackage{algorithm}
\usepackage[noend]{algpseudocode}
\usepackage{varwidth}

\usepackage{subcaption}
\usepackage{tikz}
\usetikzlibrary{positioning}
\usetikzlibrary{arrows.meta,arrows}

\usepackage{comment}

\usepackage[backend=biber, sorting=none]{biblatex}
\usepackage{multirow}
\usepackage{orcidlink}
\usepackage{afterpage}

\usepackage{hyperref}

\usepackage{caption}
\begin{document}
\mainmatter

\title{Network Centrality Metrics Based on Unrestricted Paths, Walks, and Cycles Compared to Standard Centrality Metrics}

\titlerunning{Comparing Centrality Metrics}

\author{Juuso Luhtala \orcidlink{0009-0006-3584-2365} \and Vesa Kuikka \orcidlink{0000-0002-3677-816X} \and Kimmo K. Kaski \orcidlink{0000-0002-3805-9687}}

\authorrunning{Luhtala et al.}

\tocauthor{Juuso Luhtala, Vesa Kuikka, Kimmo Kaski}
\institute{Department of Computer Science\\
  Aalto University School of Science,
  P.O. Box 11000, 00076 Aalto, Finland \\
  Correspondence: \email{juuso.luhtala@aalto.fi}}

\maketitle

\begin{abstract}
Traditional measures of closeness and betweenness centrality in networks rely on the shortest paths between nodes. Many standard metrics fail to accurately reflect the physical or probabilistic characteristics of nodal centrality and network flow, often overlooking processes such as cyclic and recurrent spreading. Here, we present new metrics based on our influence spreading model. These probabilistic measures consider all feasible paths, walks, and cycles within the network. We define in-centrality to assess how central a node is as a target of influence, and out-centrality for its role as a source of influence. We compare our metrics with standard ones by analyzing node rankings, using scatter plots, and calculating the Pearson correlation and Spearman's rank correlation coefficients. Our findings show that the betweenness centrality defined by the influence spreading model emphasizes the importance of alternative routes while maintaining similarity to standard betweenness centrality.

\end{abstract}

\section{Introduction} \label{Introduction}

Network science is a rapidly growing field that investigates the structure and dynamics of interdependent phenomena in complex systems across various disciplines. It has been particularly useful in social sciences, communication networks, and biological networks. Key methodologies in this field include statistical analyses, visualization techniques, and metrics that describe network properties. A significant area of research is the identification of clusters and communities in networks, for which many detection methods have been developed.

In the analysis of networks, many methods focus on the structural properties of paths, walks, and cycles. A walk is an alternating sequence of nodes and edges, with each node connected to its preceding and following edges. A path is a specific type of walk in which all nodes and edges are distinct. Therefore, while all paths are walks, not all walks are paths. A cycle, on the other hand, is a walk in which the edges are distinct and only the first and last nodes may be repeated.

To characterize the local and global network structures, various centrality metrics have been developed~\cite{qiu2021ranking}. Standard network centrality metrics, such as degree centrality, closeness centrality, and betweenness centrality, have become common in network science. The degree centrality $C_D(v)$ of the node $v$ measures the number of neighbors $k_v$ that the node has with respect to the total possible number of neighbors~\cite{saxena2020centrality}, reading
\begin{equation} \label{eq:degree_centrality}
C_D(v) = \frac{k_v}{N-1},
\end{equation}
where $N$ is the total number of nodes in the network. The closeness centrality $C_C(v)$ of a node $v$ measures how close the node is to all other nodes in the network~\cite{saxena2020centrality}. It is calculated as the reciprocal of the average length of the shortest paths between node $v$ and all other network nodes
\begin{equation} \label{eq:closeness}
    C_C(v)= \frac{1}{\frac{\sum_{u \neq v}{d(u,v)}}{N-1}} = \frac{N - 1}{\sum_{u \neq v}{d(u,v)}}.
\end{equation}

The betweenness centrality $b(v)$ measures how frequently a node $v$ lies on the shortest path between two other nodes in the network
\begin{equation} \label{eq:standard_betweenness}
    b(v)=\sum_{\substack{s \neq v \neq t \\s \neq t}}\frac{\sigma_{st}(v)}{\sigma_{st}},
\end{equation}
where $\sigma_{st}$ represents the total number of shortest paths from node $s$ to node $t$, while $\sigma_{st}(v)$ denotes the number of paths that pass through node $v$. A high value of betweenness centrality indicates that a node acts often as a gatekeeper of information or resources. There are numerous variants of closeness and betweenness centrality measures that can be found in the literature~\cite{saxena2020centrality, freeman1991}.

Standard centrality measures have limitations, being optimal for some applications but not for others~\cite{landherr2010, bockholt2018, bockholt2021}. When selecting metrics, it is crucial to consider whether the network is directed or undirected and what aspects are under analysis, such as hubs or influential nodes. Standard network centrality measures, like closeness and betweenness centrality, focus on the shortest paths between nodes, which limits their effectiveness by ignoring other significant paths and walks. Our proposed metrics consider all possible walks up to a specified maximum length, capturing relevant paths and walks even when edge probabilities are low due to decay effects.

Centrality measures imply specific flow patterns in networks~\cite{borgatti2005}. Two main types of process occur in networks: self-avoiding walks, where nodes are visited only once, and node-state-independent walks, with no such restrictions. Social network analysis is used to identify influential leaders and trace the spread of ideas and innovations~\cite{brass2022}. It also helps to assess information dissemination processes within networks. Influence may require multiple interactions to produce opinion changes, whereas information can spread through a single interaction~\cite{shelke2019}, highlighting the need to account for different spreading mechanisms.

Modeling epidemic spread has gained importance since the COVID-19 pandemic, leading to new approaches to understanding spread and containment at the organizational level~\cite{kuikkaSciRep}. Commonly used outcome metrics include network centrality and reproduction numbers. Communication networks are assessed using metrics that indicate connectivity and reliability, with self-avoiding walk models being relevant, especially in cyber threat investigations. Detailed modeling at the enterprise and cross-enterprise levels is essential, given the complexity of attack techniques and environments~\cite{kuikka2025network}.

In the present study, we propose new metrics to describe consistent network characteristics, which are designed to be applicable for a wide range of network science scenarios~\cite{kuikkaSciRep}. These new metrics consider all feasible paths, walks, and cycles within the network structure. In Section~\ref{related_work}, we briefly discuss the research related to our study. In Section~\ref{ISM_new_metrics}, we introduce the influence spreading model, give a motivating example of how the model works, and finally introduce our new metrics. In Section~\ref{Results}, we present the results of how the new metrics behave on multiple data sets compared to standard centrality metrics. In Section~\ref{discussion}, we discuss our research results and how they relate to network modeling. Finally, in Section~\ref{conclusions} we summarize our findings.

\section{Related Work} \label{related_work}

Centrality measures based on shortest paths make simplistic assumptions of how information, traffic or other network processes flow through a network~\cite{freeman1991, borgatti2005, bockholt2021}. The spread of information can occur either randomly or intentionally take a longer and indirect route with many intermediaries~\cite{freeman1991}. Stephen Borgatti noted that a flow process can utilize unrestricted walks, and the outcome can be either transfer (no duplication), serial duplication, or parallel duplication, depending on the nature of the flow process~\cite{borgatti2005}. Bockholt and Zweig noted that the use of certain centrality measures contains an implicit assumption on how the network process flows through the network~\cite{bockholt2018}. In another publication, the authors also note that standard betweenness centrality is not the best at identifying influential nodes in an information spreading process~\cite{bockholt2021}. However, Freeman~\cite{freeman1991} and Krackhardt~\cite{krackhardt1990} argued that standard betweenness centrality is closely associated with social power, since a node with high betweenness centrality can control communication and the flow of information. Krackhardt also noted that a high standard betweenness centrality value can indicate that a node with high betweenness has access to a variety of non-redundant sources of information.

The relationship between degree centrality and closeness centrality was recently studied in~\cite{evans2022}. The authors found a non-linear relationship between degree centrality and closeness centrality. Degree, closeness, betweenness, eigenvector, and Katz centralities were mathematically examined with respect to whether these measures fulfill the authors' three basic requirements for centrality measures in~\cite{landherr2010}. The correlation between community-aware centrality measures and classical centrality measures was studied in~\cite{rajeh2021}.

A comprehensive survey of centrality measures in complex networks provides a nice summary and overview of centrality measures~\cite{saxena2020centrality}. The survey highlights how degree centrality is a centrality measure based on local information, whereas closeness centrality and betweenness centrality make use of global knowledge of the network. Bovet and Makse give an overview of centrality measures based on whether they are based on shortest paths, walks or random walks~\cite{bovet2022centralities}. They note that degree centrality can actually be understood as the number of walks of length one starting from a node.

\section{Influence Spreading Model and New Centrality Metrics} \label{ISM_new_metrics}

In this section, we will introduce the influence spreading model, present a motivating example of how it works, and introduce new metrics defined consistently by the influence spreading model.

\subsection{Influence Spreading Model} \label{ISM}

Our influence spreading model, previously published in~\cite{kuikkaSciRep}, provides a detailed description of the network structure and non-conserving probabilistic influence flow in  that structure. The edges in the network can be directed or bidirectional, and both nodes and edges are assigned individual weights or probabilities to transmit influence. The model supports both circular and recurrent spreading throughout the network. In addition, a version of the model has been developed to implement only self-avoiding walks, meaning that each node can appear only once on a given walk. As the model can account for breakthrough effects, it enables us to describe scenarios that lie between these two extreme cases.

The influence spreading model allows for two distinct types of spreading, namely complex contagion and simple contagion. Complex contagion allows unrestricted self-intersecting walks and cycles, with the only limitations for walks being a specified maximum walk length and that the target node cannot self-intersect. However, simple contagion does not allow for self-intersecting walks or cycles. In complex contagion, the focus is on the first arrival or the first influence on a target node. Before reaching the target node, the walks are limited by the specified maximum length. The calculation for each walk ends when it reaches the target node. In this article, we focus only on complex contagion since we wish to study centrality measures related to unrestricted paths, walks, and cycles and compare them to standard centrality measures.

The probability of spreading from node $s$ to node $t$ is calculated by considering all possible walks that are shorter than or equal to a specified maximum allowed walk length $L_{max}$. We assume that the spreading process starts with probability one from the initial node $s$. It is also important to emphasize that the influence spreading model is an analytical model and not a simulation model. Due to the analytical nature of the model, its probability outputs are exact within the limitations of floating-point number calculations.

The complete logic for calculating the probability of spreading the influence was originally presented in~\cite{kuikka2022efficiency}. Depending on the context, the original article often used the term path interchangeably to mean either a path or a walk as defined in Section \ref{Introduction}. In this article, we will aim to use more exact terminology that aligns with existing graph theory terminology.

The full influence spreading model specifies a time-dependent component in addition to edge and node probabilities~\cite{kuikka2022efficiency}. In the full model, this time-dependent probability component approaches one as time grows unbounded. In this article, we assume that the probability of the time component is equal to one. By setting the time component probability to one, we can have a clearer focus on how the influence spreading model's centrality measures behave as the edge probability is varied.

In~\cite{kuikka2022efficiency}, a detailed algorithmic description was given on how different walks and their probabilities are combined. Next, we will explain the logic of the combination process of walks and the algorithm. Understanding how the influence spreading probability is calculated when walks are only restricted by the maximum allowed length of the walk is crucial to understanding the results that are presented in Section \ref{Results}.

The probabilities of two walks $\mathcal{L}_1$ and $\mathcal{L}_2$ are combined by using their longest common prefix (LCP) walk $\mathcal{L}_3$ and the rules of probability theory. The reason behind this is to ensure that the results can be interpreted as valid probabilities. If $\mathcal{L}_1$ and $\mathcal{L}_2$ do not share any common walk, then we set $\mathbb{P}(\mathcal{L}_3) = 1$, denoting the probability with $\mathbb{P}$. The influence spreading model assumes that two walks $\mathcal{L}_1$ and $\mathcal{L}_2$ are conditionally independent given a common prefix walk $\mathcal{L}_3$ i.e.\ $\mathbb{P}(\mathcal{L}_1 \cap \mathcal{L}_2 \mid \mathcal{L}_3) = \mathbb{P}(\mathcal{L}_1 \mid \mathcal{L}_3) \mathbb{P}(\mathcal{L}_2 \mid \mathcal{L}_3)$. The formula for combining the probabilities of the walks $\mathcal{L}_1$ and $\mathcal{L}_2$ given a common prefix walk $\mathcal{L}_3$ is derived as follows
\begin{align}
\mathbb{P}(\mathcal{L}_1 \cup \mathcal{L}_2) &= \mathbb{P}(\mathcal{L}_1) + \mathbb{P}(\mathcal{L}_2) - \mathbb{P}(\mathcal{L}_1 \cap \mathcal{L}_2) \notag = \mathbb{P}(\mathcal{L}_1) +  \mathbb{P}(\mathcal{L}_2) - \mathbb{P}(\mathcal{L}_1 \cap \mathcal{L}_2 \cap \mathcal{L}_3) \notag \\
&= \mathbb{P}(\mathcal{L}_1) + \mathbb{P}(\mathcal{L}_2) - \mathbb{P}(\mathcal{L}_3) \mathbb{P}(\mathcal{L}_1 \cap \mathcal{L}_2 \mid \mathcal{L}_3) \notag \\
&= \mathbb{P}(\mathcal{L}_1) + \mathbb{P}(\mathcal{L}_2) - \mathbb{P}(\mathcal{L}_3) \mathbb{P}(\mathcal{L}_1 \mid \mathcal{L}_3) \mathbb{P}(\mathcal{L}_2 \mid \mathcal{L}_3) \notag \\
&= \mathbb{P}(\mathcal{L}_1) + \mathbb{P}(\mathcal{L}_2) - \mathbb{P}(\mathcal{L}_3) \frac{\mathbb{P}(\mathcal{L}_1)}{\mathbb{P}(\mathcal{L}_3)} \frac{\mathbb{P}(\mathcal{L}_2)}{\mathbb{P}(\mathcal{L}_3)} \notag \\
&= \mathbb{P}(\mathcal{L}_1) + \mathbb{P}(\mathcal{L}_2) - \frac{\mathbb{P}(\mathcal{L}_1) \mathbb{P}(\mathcal{L}_2)}{\mathbb{P}(\mathcal{L}_3)} \label{path_combination},
\end{align} where the second equality follows from the walks $\mathcal{L}_1$ and $\mathcal{L}_2$ both sharing the common walk $\mathcal{L}_3$, the third equality follows from the definition of conditional probability, the fourth equality follows from assumed conditional independence, and the fifth equality follows from $\mathbb{P}(\mathcal{L}_i) = \mathbb{P}(\mathcal{L}_i \cap \mathcal{L}_3) = \mathbb{P}(\mathcal{L}_3) \mathbb{P}(\mathcal{L}_i \mid \mathcal{L}_3)$ for $i = 1, 2$.

Equation (\ref{path_combination}) for combining two walks $\mathcal{L}_1$ and $\mathcal{L}_2$ forms the basis of the influence spreading algorithm \ref{alg:algorithm1}. The algorithm \ref{alg:algorithm1} and a more efficient algorithm for calculating influence spreading were originally presented in~\cite{kuikka2022efficiency}. For ease of understanding, we present the version that explicitly lists all possible allowed walks, although it is computationally less efficient. The computational limitations of algorithm \ref{alg:algorithm1} were discussed at length in~\cite{kuikka2022efficiency}. In practice, we have used the more efficient algorithm in our experiments. It is important to emphasize that both algorithm versions produce the exact same results.
\begin{algorithm}
\caption{Influence Spreading Algorithm} \label{alg:algorithm1}
\begin{algorithmic}[1]
\Procedure{Spread}{$s, t, G$}
\State $\mathcal{K} \gets \text{list of all walks from node } s \text{ to  node } t \text{ in } G \text{ with at most } L_{max} \text{ edges}$
\If {$\mathcal{K} = \emptyset$} 
  \Return 0
\EndIf 
\State $\text{Sort } \mathcal{K} \text{ in lexicographically increasing order based on walk prefixes}$
\For {$c = L_{max} \dots 0$}
  \If {$\text{there are no walks of at least } c \text{ edges}$}
    \State $\textbf{continue}$
  \EndIf
  \While {$\text{there are walks } \mathcal{L}_1 \text{ and } \mathcal{L}_2 \text{ in } \mathcal{K} \text{ with an LCP walk } \mathcal{L}_3 \text{ of }  c \text{ edges}$}
    \State $\mathbb{P}(\mathcal{L}_1) \gets \mathbb{P}(\mathcal{L}_1) + \mathbb{P}(\mathcal{L}_2) - \mathbb{P}(\mathcal{L}_1) \mathbb{P}(\mathcal{L}_2) / \mathbb{P}(\mathcal{L}_3)$ \Comment{Equation (\ref{path_combination})}
    \State $\text{Remove } \mathcal{L}_2 \text{ from } \mathcal{K}$
  \EndWhile
\EndFor
\State \Return $\mathbb{P}(\mathcal{L}_1)$ \Comment{The probability of the only walk left in $\mathcal{K}$}
\EndProcedure
\end{algorithmic}
\end{algorithm}

The influence spreading algorithm works by exploring all possible walks from source node $s$ to target node $t$ with at most $L_{max}$ edges. The possible walks are then sorted in increasing lexicographic order (shorter walks first; walks with the same length are sorted according to where they first differ). The walk combination process is done by first combining the walks with the longest common prefix (LCP). Then shorter LCPs are considered until all walks have been combined into one. Walks having an LCP of the same length can be combined in any order. In probabilistic terms, the actual combination of walk probabilities is calculated using equation (\ref{path_combination}).

\subsection{Motivating Example of the Influence Spreading Model} \label{illustrative_example}

We will now present an illustrative and motivating example of how the influence spreading algorithm works. This example will not only explain the influence spreading model, but also give insight into the results presented in Section \ref{Results}.

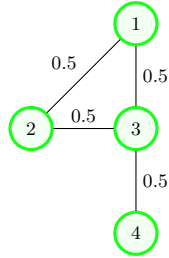
\begin{figure}
\centering
\begin{tikzpicture}[scale=0.8, transform shape, roundnode/.style={circle, draw=green!90, fill=green!5, very thick, minimum size=7mm},]
\node[roundnode] (3) {3};
\node[roundnode] (1) [above=of 3] {1};
\node[roundnode] (2) [left=of 3] {2};
\node[roundnode] (4) [below=of 3] {4};

\draw[-] (1) -- (3) node[midway, right] {0.5};
\draw[-] (1) -- (2) node[midway, above left] {0.5};
\draw[-] (2) -- (3) node[midway, above] {0.5};
\draw[-] (3) -- (4) node[midway, right] {0.5};
\end{tikzpicture}
\caption{A small undirected network where the edge weights are probabilities.} \label{fig:simplest_example}
\end{figure}

Consider an undirected four-node network presented in Figure \ref{fig:simplest_example}, where edge weights denote influencing probabilities. While the network in Figure \ref{fig:simplest_example} is small, it shows some key features of the influence spreading model. Here we calculate the influence spreading probabilities for the maximum walk length of $L_{max} = 5$. We consider two separate examples, first using node $1$ as the source and node $4$ as the target, and secondly using node $4$ as the source and node $1$ as the target. These calculations result in different end probabilities since there are more cycle possibilities when going from node $1$ to node $4$ than vice versa.

When node $1$ is the source and node $4$ is the target, we have in total ten different walks that have length 5 or less. These ten walks are illustrated in Figure \ref{fig:paths_from_1_to_4}. In contrast, when node $4$ is the source and node $1$ is the target, then we have in total only six different walks that have length 5 or less. These six walks are illustrated in Figure \ref{fig:paths_from_4_to_1}.
\begin{figure}
\begin{subfigure}{1.0 \textwidth}
\begin{subfigure}{0.18 \textwidth}
\begin{tikzpicture}[scale=0.9, transform shape, roundnode/.style={circle, draw=green!90, fill=green!5, very thick},]
\node[roundnode] (3) {3};
\node[roundnode] (1) [above=of 3] {1};
\node[roundnode] (4) [below=of 3] {4};
\draw[-{Stealth[scale=1.5]}] (1) -- (3) node[midway, left] {1.};
\draw[-{Stealth[scale=1.5]}] (3) -- (4) node[midway, left] {2.};
\end{tikzpicture}
\end{subfigure}
\begin{subfigure}{0.18 \textwidth}
\begin{tikzpicture}[scale=0.9, transform shape, roundnode/.style={circle, draw=green!90, fill=green!5, very thick},]
\node[roundnode] (3) {3};
\node[roundnode] (1) [above=of 3] {1};
\node[roundnode] (2) [left=of 3] {2};
\node[roundnode] (4) [below=of 3] {4};
\draw[-{Stealth[scale=1.5]}] (1) -- (2) node[midway, left] {1.};
\draw[-{Stealth[scale=1.5]}] (2) -- (3) node[midway, below] {2.};
\draw[-{Stealth[scale=1.5]}] (3) -- (4) node[midway, left] {3.};
\end{tikzpicture}
\end{subfigure}
\begin{subfigure}{0.23 \textwidth}
\begin{tikzpicture}[scale=0.9, transform shape, roundnode/.style={circle, draw=green!90, fill=green!5, very thick},]
\node[roundnode] (3) {3};
\node[roundnode] (1) [above=of 3] {1};
\node[roundnode] (2) [left=of 3] {2};
\node[roundnode] (4) [below=of 3] {4};
\draw[-{Stealth[scale=1.5]}] (1) -- (2) node[midway, left] {1.};
\draw[-{Stealth[scale=1.5]}] (2) .. controls +(up:1cm) and +(left:1cm) .. (1) node[midway, above] {2.};
\draw[-{Stealth[scale=1.5]}] (1) -- (3) node[midway, right] {3.};
\draw[-{Stealth[scale=1.5]}] (3) -- (4) node[midway, left] {4.};
\end{tikzpicture}
\end{subfigure}
\begin{subfigure}{0.18 \textwidth}
\begin{tikzpicture}[scale=0.9, transform shape, roundnode/.style={circle, draw=green!90, fill=green!5, very thick},]
\node[roundnode] (3) {3};
\node[roundnode] (1) [above=of 3] {1};
\node[roundnode] (4) [below=of 3] {4};
\draw[-{Stealth[scale=1.5]}] (1) .. controls +(left:1cm) and +(left:1cm) .. (3) node[midway, left] {1.};
\draw[-{Stealth[scale=1.5]}] (3) -- (1) node[midway, left] {2.};
\draw[-{Stealth[scale=1.5]}] (1) .. controls +(right:1cm) and +(right:1cm) .. (3) node[near start, right] {3.};
\draw[-{Stealth[scale=1.5]}] (3) -- (4) node[midway, left] {4.};
\end{tikzpicture}
\end{subfigure}
\begin{subfigure}{0.18 \textwidth}
\begin{tikzpicture}[scale=0.9, transform shape, roundnode/.style={circle, draw=green!90, fill=green!5, very thick},]
\node[roundnode] (3) {3};
\node[roundnode] (1) [above=of 3] {1};
\node[roundnode] (2) [left=of 3] {2};
\node[roundnode] (4) [below=of 3] {4};
\draw[-{Stealth[scale=1.5]}] (1) -- (3) node[midway, right] {1.};
\draw[-{Stealth[scale=1.5]}] (3) .. controls +(up:0.4cm) and +(up:1cm) .. (2) node[midway, above] {2.};
\draw[-{Stealth[scale=1.5]}] (2) -- (3) node[midway, below] {3.};
\draw[-{Stealth[scale=1.5]}] (3) -- (4) node[midway, left] {4.};
\end{tikzpicture}
\end{subfigure}
\par\bigskip
\begin{subfigure}{0.18 \textwidth}
\begin{tikzpicture}[scale=0.9, transform shape, roundnode/.style={circle, draw=green!90, fill=green!5, very thick},]
\node[roundnode] (3) {3};
\node[roundnode] (1) [above=of 3] {1};
\node[roundnode] (2) [left=of 3] {2};
\node[roundnode] (4) [below=of 3] {4};
\draw[-{Stealth[scale=1.5]}] (1) .. controls +(left:1cm) and +(up:1cm) .. (2) node[midway, above] {1.};
\draw[-{Stealth[scale=1.5]}] (2) -- (1) node[midway, left] {2.};
\draw[-{Stealth[scale=1.5]}] (1) .. controls +(down:1cm) .. (2) node[midway, above] {3.};
\draw[-{Stealth[scale=1.5]}] (2) -- (3) node[midway, below] {4.};
\draw[-{Stealth[scale=1.5]}] (3) -- (4) node[midway, left] {5.};
\end{tikzpicture}
\end{subfigure}
\begin{subfigure}{0.18 \textwidth}
\begin{tikzpicture}[scale=0.9, transform shape, roundnode/.style={circle, draw=green!90, fill=green!5, very thick},]
\node[roundnode] (3) {3};
\node[roundnode] (1) [above=of 3] {1};
\node[roundnode] (2) [left=of 3] {2};
\node[roundnode] (4) [below=of 3] {4};
\draw[-{Stealth[scale=1.5]}] (1) -- (2) node[midway, above] {1.};
\draw[-{Stealth[scale=1.5]}] (2) -- (3) node[midway, below] {2.};
\draw[-{Stealth[scale=1.5]}] (3) .. controls +(right:0.2cm) and +(right:1cm) .. (1) node[near end, right] {3.};
\draw[-{Stealth[scale=1.5]}] (1) -- (3) node[midway, left] {4.};
\draw[-{Stealth[scale=1.5]}] (3) -- (4) node[midway, left] {5.};
\end{tikzpicture}
\end{subfigure}
\begin{subfigure}{0.18 \textwidth}
\begin{tikzpicture}[scale=0.9, transform shape, roundnode/.style={circle, draw=green!90, fill=green!5, very thick},]
\node[roundnode] (3) {3};
\node[roundnode] (1) [above=of 3] {1};
\node[roundnode] (2) [left=of 3] {2};
\node[roundnode] (4) [below=of 3] {4};
\draw[-{Stealth[scale=1.5]}] (1) -- (2) node[near start, above] {1.};
\draw[-{Stealth[scale=1.5]}] (2) .. controls +(up:1cm) and +(up:1cm) .. (3) node[near start, above] {2.};
\draw[-{Stealth[scale=1.5]}] (3) -- (2) node[midway, above] {3.};
\draw[-{Stealth[scale=1.5]}] (2) .. controls +(down:1cm) .. (3) node[midway, below] {4.};
\draw[-{Stealth[scale=1.5]}] (3) -- (4) node[midway, right] {5.};
\end{tikzpicture}
\end{subfigure}
\begin{subfigure}{0.18 \textwidth}
\begin{tikzpicture}[scale=0.9, transform shape, roundnode/.style={circle, draw=green!90, fill=green!5, very thick},]
\node[roundnode] (3) {3};
\node[roundnode] (1) [above=of 3] {1};
\node[roundnode] (2) [left=of 3] {2};
\node[roundnode] (4) [below=of 3] {4};
\draw[-{Stealth[scale=1.5]}] (1) -- (3) node[midway, left] {1.};
\draw[-{Stealth[scale=1.5]}] (3) .. controls +(up:0.1cm) and +(right:1cm) .. (1) node[midway, right] {2.};
\draw[-{Stealth[scale=1.5]}] (1) -- (2) node[midway, above] {3.};
\draw[-{Stealth[scale=1.5]}] (2) -- (3) node[midway, below] {4.};
\draw[-{Stealth[scale=1.5]}] (3) -- (4) node[midway, left] {5.};
\end{tikzpicture}
\end{subfigure}
\begin{subfigure}{0.18 \textwidth}
\begin{tikzpicture}[scale=0.9, transform shape, roundnode/.style={circle, draw=green!90, fill=green!5, very thick},]
\node[roundnode] (3) {3};
\node[roundnode] (1) [above=of 3] {1};
\node[roundnode] (2) [left=of 3] {2};
\node[roundnode] (4) [below=of 3] {4};
\draw[-{Stealth[scale=1.5]}] (1) -- (3) node[midway, left] {1.};
\draw[-{Stealth[scale=1.5]}] (3) -- (2) node[midway, above] {2.};
\draw[-{Stealth[scale=1.5]}] (2) -- (1) node[midway, above] {3.};
\draw[-{Stealth[scale=1.5]}] (1) .. controls +(right:1cm) and +(right:1cm) .. (3) node[midway, right] {4.};
\draw[-{Stealth[scale=1.5]}] (3) -- (4) node[midway, right] {5.};
\end{tikzpicture}
\end{subfigure}
\caption{All ten walks from node 1 to node 4 with walk length equal or less than $5$.} 
\vspace{+12pt}
\label{fig:paths_from_1_to_4}
\end{subfigure}

\begin{subfigure}{1.0 \textwidth}
\begin{subfigure}{0.18 \textwidth}
\begin{tikzpicture}[scale=0.9, transform shape, roundnode/.style={circle, draw=green!90, fill=green!5, very thick},]
\node[roundnode] (3) {3};
\node[roundnode] (1) [above=of 3] {1};
\node[roundnode] (4) [below=of 3] {4};
\draw[-{Stealth[scale=1.5]}] (4) -- (3) node[midway, left] {1.};
\draw[-{Stealth[scale=1.5]}] (3) -- (1) node[midway, left] {2.};
\end{tikzpicture}
\end{subfigure}
\begin{subfigure}{0.18 \textwidth}
\begin{tikzpicture}[scale=0.9, transform shape, roundnode/.style={circle, draw=green!90, fill=green!5, very thick},]
\node[roundnode] (3) {3};
\node[roundnode] (1) [above=of 3] {1};
\node[roundnode] (2) [left=of 3] {2};
\node[roundnode] (4) [below=of 3] {4};
\draw[-{Stealth[scale=1.5]}] (4) -- (3) node[midway, left] {1.};
\draw[-{Stealth[scale=1.5]}] (3) -- (2) node[midway, below] {2.};
\draw[-{Stealth[scale=1.5]}] (2) -- (1) node[midway, left] {3.};
\end{tikzpicture}
\end{subfigure}
\begin{subfigure}{0.18 \textwidth}
\begin{tikzpicture}[scale=0.9, transform shape, roundnode/.style={circle, draw=green!90, fill=green!5, very thick},]
\node[roundnode] (3) {3};
\node[roundnode] (1) [above=of 3] {1};
\node[roundnode] (2) [left=of 3] {2};
\node[roundnode] (4) [below=of 3] {4};
\draw[-{Stealth[scale=1.5]}] (4) -- (3) node[near start, left] {1.};
\draw[-{Stealth[scale=1.5]}] (3) -- (2) node[midway, above] {2.};
\draw[-{Stealth[scale=1.5]}] (2) .. controls +(down:1cm) .. (3) node[midway, below] {3.};
\draw[-{Stealth[scale=1.5]}] (3) -- (1) node[midway, right] {4.};
\end{tikzpicture}
\end{subfigure}
\begin{subfigure}{0.18 \textwidth}
\begin{tikzpicture}[scale=0.9, transform shape, roundnode/.style={circle, draw=green!90, fill=green!5, very thick},]
\node[roundnode] (3) {3};
\node[roundnode] (1) [above=of 3] {1};
\node[roundnode] (4) [below=of 3] {4};
\draw[-{Stealth[scale=1.5]}] (4) .. controls +(left:1cm) and +(left:1cm) .. (3) node[near start ,left] {1.};
\draw[-{Stealth[scale=1.5]}] (3) -- (4) node[midway, left] {2.};
\draw[-{Stealth[scale=1.5]}] (4) .. controls +(right:1cm) .. (3) node[near end, right] {3.};
\draw[-{Stealth[scale=1.5]}] (3) -- (1) node[midway, right] {4.};
\end{tikzpicture}
\end{subfigure}
\begin{subfigure}{0.18 \textwidth}
\begin{tikzpicture}[scale=0.9, transform shape, roundnode/.style={circle, draw=green!90, fill=green!5, very thick},]
\node[roundnode] (3) {3};
\node[roundnode] (1) [above=of 3] {1};
\node[roundnode] (2) [left=of 3] {2};
\node[roundnode] (4) [below=of 3] {4};
\draw[-{Stealth[scale=1.5]}] (4) -- (3) node[near start, right] {1.};
\draw[-{Stealth[scale=1.5]}] (3) .. controls +(down:0.2cm) and +(down:1cm) .. (2) node[near end, below] {2.};
\draw[-{Stealth[scale=1.5]}] (2) -- (3) node[midway, below] {3.};
\draw[-{Stealth[scale=1.5]}] (3) .. controls +(up:1cm) .. (2) node[near start, right] {4.};
\draw[-{Stealth[scale=1.5]}] (2) -- (1) node[midway, left] {5.};
\end{tikzpicture}
\end{subfigure}
\par\bigskip
\begin{subfigure}{0.18 \textwidth}
\begin{tikzpicture}[scale=0.9, transform shape, roundnode/.style={circle, draw=green!90, fill=green!5, very thick},]
\node[roundnode] (3) {3};
\node[roundnode] (1) [above=of 3] {1};
\node[roundnode] (2) [left=of 3] {2};
\node[roundnode] (4) [below=of 3] {4};
\draw[-{Stealth[scale=1.5]}] (4) .. controls +(left:1cm) .. (3) node[near start, left] {1.};
\draw[-{Stealth[scale=1.5]}] (3) -- (4) node[midway, right] {2.};
\draw[-{Stealth[scale=1.5]}] (4) .. controls +(right:1cm) and +(right:1cm) .. (3) node[midway, right] {3.};
\draw[-{Stealth[scale=1.5]}] (3) -- (2) node[midway, below] {4.};
\draw[-{Stealth[scale=1.5]}] (2) -- (1) node[midway, left] {5.};
\end{tikzpicture}
\end{subfigure}
\vspace{+2pt}
\caption{All six walks from node 4 to node 1 with walk length equal to or less than $5$.} \label{fig:paths_from_4_to_1}
\end{subfigure}
\vspace{+2pt}
\caption{Example walks with walk length equal or less than $5$ for the network depicted in Figure \ref{fig:simplest_example}. The numbers denote the order of traversal along the edges of the network.}
\end{figure}
When the influence spreading algorithm \ref{alg:algorithm1} is performed with node 4 as the source and node 1 as the target, we find six walks with the walk length shorter than or equal to $5$. The walks are then sorted in increasing lexicographic order as follows
(i) $4 \rightarrow 3 \rightarrow 1$, (ii) $4 \rightarrow 3 \rightarrow 2 \rightarrow 1$, (iii) $4 \rightarrow 3 \rightarrow 2 \rightarrow 3 \rightarrow 1$, (iv) $4 \rightarrow 3 \rightarrow 4 \rightarrow 3 \rightarrow 1$, (v) $4 \rightarrow 3 \rightarrow 2 \rightarrow 3 \rightarrow 2 \rightarrow 1$, and (vi) $4 \rightarrow 3 \rightarrow 4 \rightarrow 3 \rightarrow 2 \rightarrow 1$.
The longest common prefix (LCP) walks in this lexicographic order are walks including $4 \rightarrow 3 \rightarrow 2 \rightarrow3$ and $4 \rightarrow 3 \rightarrow 4 \rightarrow 3$. We can process walks with the same LCP walk lengths in any order. An example calculation of the combination of walk probabilities is combining the third $4 \rightarrow 3 \rightarrow 2 \rightarrow 3 \rightarrow 1$ and fifth $4 \rightarrow 3 \rightarrow 2 \rightarrow 3 \rightarrow 2 \rightarrow 1$ walks. The third walk has probability $0.5^4$, the fifth walk has probability $0.5^5$, and their LCP walk has probability $0.5^3$. Therefore, the combined probability is $0.5^4 + 0.5^5 - 0.5^4 \cdot 0.5^5 / 0.5^3 = 0.078125$. When the algorithm \ref{alg:algorithm1} is executed until completion, we get the probability of approximately $0.358$. This means that node 4 influences node 1 with an approximate probability of $0.358$. When the influence spreading algorithm \ref{alg:algorithm1} is performed with node 1 as the source and node 4 as the target, we get the result that node 1 influences node 4 with approximately $0.472$ probability. Hence, we see that node 1 has more influence on node 4 than vice versa. This result is explained by the different number of walk possibilities.

Figures \ref{fig:paths_from_1_to_4} and \ref{fig:paths_from_4_to_1} underscore a crucial aspect of the influence spreading model: The calculations are conducted only until the target node is reached on each walk. This feature exists because the influence spreading model focuses on the first arrival/influence. This feature results in a different number of possible walks when going from node 1 to node 4 and vice versa. There are four fewer paths from node 4 to 1 than there are from node 1 to 4. The position of node 1 gives it more possibilities for cycles before reaching the target node 4. In contrast, the position of node 4 allows for fewer cycle possibilities in the network before reaching target node 1. We also see that node 4 is on the periphery of this small network, while node 1 is more central. This is also an important observation when we discuss the results of other experiments in Section \ref{Results}, where we will also refer to a more formalized definition of the periphery.

\subsection{New Centrality Metrics} \label{Metrics}

The output of the influence spreading model is represented as a matrix, known as the influence spreading matrix. The elements of this matrix describe the directed probabilities of influence spreading from a source node $i$ to a target node $j$ for all pairs of $N \times N$ nodes within a network consisting of $N$ nodes. The elements of the matrix reflect the probabilities of directed spreading or connectivity between nodes in the network $V$, and it is used to analyze the structure and flow processes of the network. Importantly, models other than our influence spreading model can also create this probability matrix. Using the influence spreading matrix $M$, we define our new out-centrality $C_{out}$ and in-centrality $C_{in}$ as follows
\begin{equation} \label{eq:out-centrality}
    C_{out}(i)=\frac{1}{N-1}\sum_{j \in V \setminus \{ i \}} M(i,j), \quad i=1,...N
\end{equation}
and
\begin{equation} \label{eq:in-centrality}
    C_{in}(j)=\frac{1}{N-1}\sum_{i \in V \setminus \{ j \}} M(i,j), \quad j=1,...,N.
\end{equation}

We assume that the spreading process begins with the probability of one from the initial node. When spreading occurs from branching paths or walks, it can happen in multiple directions, indicating the likelihood of first arrival at the end nodes. By ignoring the diagonal elements of the influence spreading matrix (a node's influence on itself is always one), we can define the out-centrality and in-centrality values. To normalize these values to one (or express them as percentages), we use the factor $1/(N-1)$ on the right side of equations (\ref{eq:out-centrality}) and (\ref{eq:in-centrality}). Out-centrality measures how much a node influences other nodes on average, and in-centrality measures how much a node is influenced by other nodes on average.

The influence spreading model also defines its own betweenness centrality measure. We refer to this betweenness as ISM betweenness centrality or ISM betweenness, which is defined with the help of the concept of cohesion. The cohesion $\mathcal{C}$ of the network $G = (V, E)$, where $V$ is the set of nodes and $E$ is the set of edges in the network, is defined as $$\mathcal{C} = \sum_{\substack{s, e \in V \\ s \neq e}} M(s, e).$$ The cohesion $\mathcal{C}_v$ of the network without node $v$ and its removed incident edges is defined as $$\mathcal{C}_v = \sum_{\substack{s, e \in V \setminus \{ v \} \\ s \neq e}} M^{*}(s, e),$$ where $M^{*}$ is the influence spreading matrix calculated from the network without node $v$ and its incident edges. With these definitions, we define ISM betweenness centrality $b_{ISM}(v)$ of a node $v$ as
\begin{equation} \label{eq:ISM_betweenness}
b_{ISM}(v) = \frac{\mathcal{C} - \mathcal{C}_v}{\mathcal{C}}.
\end{equation}

It is important to emphasize the way the ISM betweenness $b_{ISM}$ is calculated. When the ISM betweenness value of node $v$ is calculated, we remove the node and its incident edges from the network, and then the logic of the influence spreading algorithm \ref{alg:algorithm1} is performed on this altered network. It is important to note that ISM betweenness measures the relative importance of node $v$ by examining how much the cohesion $\mathcal{C}$ of the network changes when node $v$ is removed. If node $v$ acts as a particularly important bridge in the network, its removal can even lead to the network being disconnected into two separate components. 

\section{Results} \label{Results}

In this section, we will compare the metrics defined by the influence spreading model to standard centrality metrics. Furthermore, we will further examine the behavior of the centrality metrics defined by the influence spreading model.
The influence spreading model defines its own betweenness centrality (\ref{eq:ISM_betweenness}) and also new centrality metrics, i.e., out-centrality (\ref{eq:out-centrality}) and in-centrality (\ref{eq:in-centrality}). The compared standard centrality metrics are betweenness centrality (\ref{eq:standard_betweenness}), degree centrality (\ref{eq:degree_centrality}) and closeness centrality (\ref{eq:closeness}).

The influence spreading model uses probabilities as edge weights. In our experiments, the edge probability is varied between experiments, but all edges of the network always have the same probability. However, the standard betweenness centrality and the closeness centrality require a distance to be defined between the nodes. We interpret the shortest path between the nodes as the path with the fewest edges (as if the distance between the nodes were one). If edge probabilities are not the same for all edge weights, then a systematic way of mapping probabilities to distances is needed. Appendix \ref{appendix} explores one possible way of mapping edge probabilities to edge distances. However, in our experiments, the shortest paths are calculated with the assumption that all edge distances are equal. Therefore, calculating the shortest path between nodes reduces to finding the path with the fewest edges.

We frequently characterize nodes as peripheral or central. Borgatti and Everett proposed formalizations for the concept of core/peripheral structure in networks~\cite{borgatti2000}. One of the formalizations they proposed was a discrete model consisting of a core and a periphery. The core was defined as a cohesive subgraph in which nodes are connected to each other in a maximal sense, and the periphery consists of nodes that are loosely connected to the core but do not share the maximal cohesion with the core. Yanchenko and Sengupta explain the core as consisting of nodes that are densely connected to each other, and some of them also have connections to peripheral nodes, whereas peripheral nodes are only sparsely connected to each other~\cite{yanchenko2023}. We use the term peripheral nodes in the sense and meaning given by Yanchenko and Sengupta, and central nodes are interpreted as belonging to the core.

In Section \ref{krackhardt}, we will have a detailed individual node-level examination of the behavior of the influence spreading model and its centrality metrics on a small network. Using a small network allows us to examine and explain the behavior of the influence spreading model and its centrality metrics with node-level detail. In Section \ref{ism_random}, we examine the collective network-level behavior of the influence spreading model and its centrality metrics on larger $1000$ node random graphs/networks. All experiments in Sections \ref{krackhardt}, and \ref{ism_random} use the maximum allowed walk length $L_{max} = 20$. In Section \ref{discussion}, the effect of using shorter or longer $L_{max}$ is discussed.

\subsection{Detailed Node-Level Analysis of the Influence Spreading Model on the Krackhardt Kite Network} \label{krackhardt}

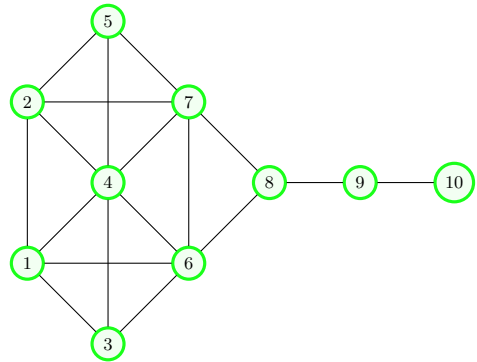
\begin{figure}
\centering
\begin{tikzpicture}[scale=0.75, transform shape, roundnode/.style={circle, draw=green!90, fill=green!5, very thick},]
\node[roundnode] (1) {1};
\node[roundnode] (3) [below right=of 1] {3};
\node[roundnode] (4) [above right=of 1] {4};
\node[roundnode] (2) [above left=of 4] {2};
\node[roundnode] (5) [above right=of 2] {5};
\node[roundnode] (6) [above right=of 3] {6};
\node[roundnode] (7) [above right=of 4] {7};
\node[roundnode] (8) [above right=of 6] {8};
\node[roundnode] (9) [right=of 8] {9};
\node[roundnode] (10) [right=of 9] {10};
\draw (1) -- (2);
\draw (1) -- (3);
\draw (1) -- (4);
\draw (1) -- (6);
\draw (2) -- (4);
\draw (2) -- (5);
\draw (2) -- (7);
\draw (3) -- (4);
\draw (3) -- (6);
\draw (4) -- (5);
\draw (4) -- (6);
\draw (4) -- (7);
\draw (5) -- (7);
\draw (6) -- (7);
\draw (6) -- (8);
\draw (7) -- (8);
\draw (8) -- (9);
\draw (9) -- (10);
\end{tikzpicture}
\caption{Krackhardt kite. This network was originally presented in~\cite{krackhardt1990}.} \label{fig:krackhardt_kite}
\end{figure}

We will examine the behavior of the influence spreading model in the Krackhardt kite network~\cite{krackhardt1990}. The Krackhardt kite shown in Figure \ref{fig:krackhardt_kite} is a useful example to compare the influence spreading model against standard centrality metrics, as it reveals the differences of the influence spreading model's centrality measures compared to standard centrality measures. The Krackhardt kite network is well known for producing a different highest ranked node for each of the calculations of standard betweenness (node 8), closeness (nodes 6 and 7), and degree centrality (node 4)~\cite{krackhardt1990}.

\begin{figure} 
\includegraphics[scale=0.35]{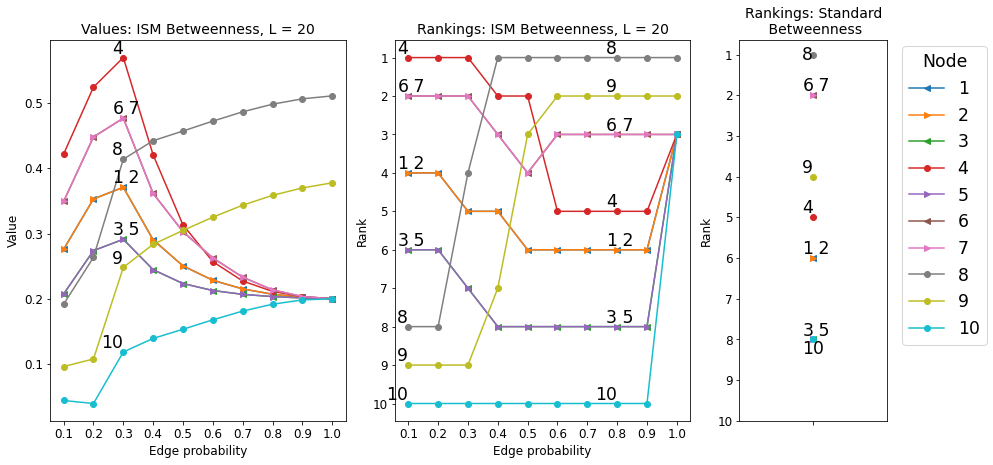} 
\caption{Influence spreading model betweenness values and rankings alongside standard betweenness centrality rankings for the Krackhardt kite network displayed in Figure \ref{fig:krackhardt_kite}. Node labels are marked alongside the plotted points.} \label{fig:krackhardt_ISM_betweenness}
\end{figure}

In Figure \ref{fig:krackhardt_ISM_betweenness}, we see the values and node rankings for the betweenness centrality measure as defined by the influence spreading model. The rankings of ISM betweenness are plotted side by side with the standard betweenness centrality rankings. The ISM betweenness centrality value is dependent on the edge probability. Figure \ref{fig:krackhardt_ISM_betweenness} also shows how the rankings between nodes change as the edge probability increases. Standard betweenness gives different rankings compared to ISM betweenness for all edge probabilities. The greatest difference is how nodes 8, 9, and 4 behave. Their rankings change quite markedly when the edge probability is increased.
It is also important to note how ISM betweenness gives non-zero values for nodes 3, 5, and 10, whereas standard betweenness assigns them the value zero since they do not appear in any shortest path between nodes. However, by allowing any walks including cycles, the influence spreading model recognizes that these nodes could have more importance than mere shortest path calculations could imply.

When edge probabilities are one, the ISM betweenness values converge into three separate values. This phenomenon is explained by how the ISM betweenness is calculated. Node 8 attains the highest ISM betweenness value for large edge probability values. When node 8 and its incident edges are removed from the network, the network is divided into two separate components: one containing two nodes and the other containing seven nodes. This results in a large difference in the cohesion values $\mathcal{C}$ and $\mathcal{C}_8$, and therefore the ISM betweenness value is high. A similar phenomenon occurs when node 9 and its incident edges are removed from the network. Again, the network is divided into two components, one containing only one node and the second containing eight nodes. For all other nodes, removing them from the network does not create two separate components. Instead, the network stays connected, which allows all influences to be maximally transferred between all nodes. Therefore, for nodes 1, 2, 3, 4, 5, 6, 7, and 10, their ISM betweenness values converge to the same value.

\begin{figure} 
\includegraphics[scale=0.35]{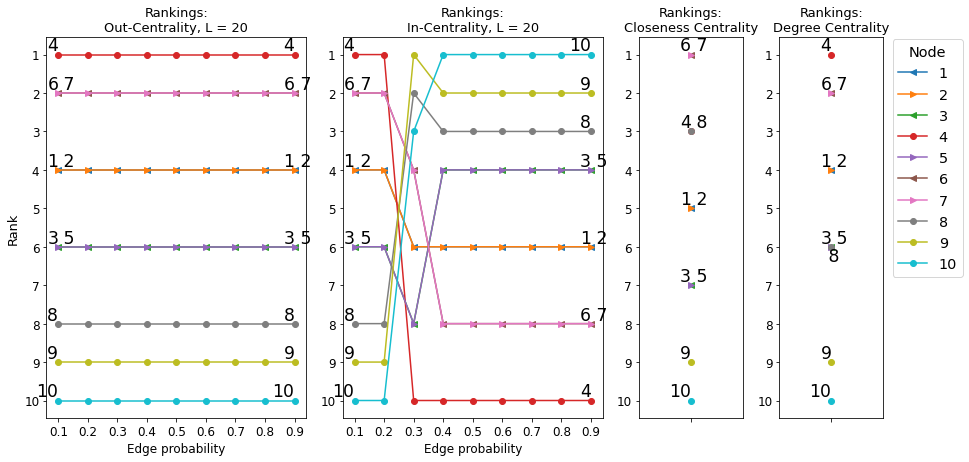} 
\caption{Influence spreading model out-centrality and in-centrality rankings compared to standard closeness and degree centrality rankings for the Krackhardt kite network displayed in Figure \ref{fig:krackhardt_kite}. Node labels are marked alongside the plotted points.} \label{fig:krackhardt_out_in_ranking_comparison}
\end{figure}

In Figure \ref{fig:krackhardt_out_in_ranking_comparison}, the node rankings given by out-centrality and in-centrality are compared to the rankings given by standard closeness and degree centrality. Since out-centrality measures a node's average influence on other nodes and in-centrality measures the other nodes' average influence on a node, these centrality metrics measure different things. Additionally, standard closeness centrality measures how close nodes are to each other and degree centrality measures the number of neighbors a node has. Therefore, all of these centrality metrics measure different things, but despite this, we notice similarities between these metrics. Out-centrality rankings remain unchanged as the edge probability increases. We see that in rankings, there is great similarity between out-centrality and standard closeness centrality, and out-centrality and degree centrality.
In Figure \ref{fig:krackhardt_out_in_ranking_comparison}, the in-centrality ranking is initially exactly the same as the out-centrality ranking. Thus, for low edge probability values, there is a great similarity to closeness centrality and degree centrality. However, as the edge probability value is increased, the in-centrality rankings start to reverse. From the edge probability value $0.4$ onward, the rankings have completely reversed. Consequently, for edge probabilities that are equal to or greater than $0.4$, there is a reverse order relationship between in-centrality and closeness centrality and between in-centrality and degree centrality.

\begin{figure} 
\includegraphics[scale=0.35]{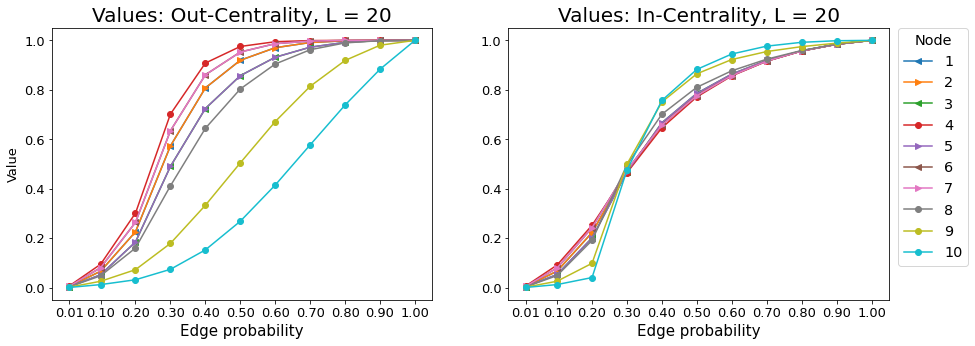} 
\caption{Influence spreading model out-centrality and in-centrality values for the Krackhardt kite network displayed in Figure \ref{fig:krackhardt_kite}. The lowest probability is $0.01$ and thereafter probabilities increase from $0.1$ to $1.0$ in $0.1$ increments.} \label{fig:krackhardt_out_in_values}
\end{figure}

In Figure \ref{fig:krackhardt_out_in_values}, we see the values for the out-centrality and in-centrality measures defined by the influence spreading model. From small to large edge probabilities, out-centrality values have much larger dispersion than in-centrality values. In Figure \ref{fig:krackhardt_out_in_ranking_comparison}, we saw how the in-centrality rankings initially developed in a similar fashion to out-centrality rankings, but for higher edge probabilities, the rankings were completely reversed. Figure \ref{fig:krackhardt_out_in_values} tells the same story with one important difference as it highlights how the value differences in in-centrality for the Krackhardt kite network are small. However, the small differences in the in-centrality values are still significant since they are explained by the network structure.

Additionally, a notable feature of the influence spreading model is also shown in Figure \ref{fig:krackhardt_out_in_values}. When edge probabilities are $1.0$ for all edges, all of the influence is coming through, and every node attains a trivial $1.0$ value for their out- and in-centrality measures. When edge probabilities are almost zero for all edges, almost none of the influence is coming through, and every node achieves a value close to zero for their out- and in-centrality measures. 
In Figures \ref{fig:krackhardt_out_in_ranking_comparison} and \ref{fig:krackhardt_out_in_values}, we see that peripheral nodes attain larger in-centrality values for high edge probabilities than more central nodes. The reason for this phenomenon is found in the example of Section \ref{illustrative_example}. When a more central node is exerting influence towards a peripheral node, it has more cycle possibilities than a peripheral node due to its position in the network. In contrast, when a more peripheral node is exerting influence towards a central node, it has fewer cycle possibilities than a central node due to its position in the network. It is also important to emphasize that if a target node has a high degree, then all of the edges leading to the target node are only allowed as the last edge on a walk. The influence spreading model focuses on the first arrival (first influence) and therefore peripheral nodes are influenced more than central nodes when edge probabilities are high, since high edge probabilities allow the influence to reach a target node even for very long walks.

Figures \ref{fig:krackhardt_out_in_ranking_comparison} and \ref{fig:krackhardt_out_in_values} demonstrate that the out-centrality ranking between nodes remains the same, although their values change as the edge probability value increases. This result can be interpreted as the edge probability acting as an amplifier to a node's position in the network. Every node can exert more influence, but the more central nodes still have more cycle possibilities than the more peripheral nodes. Therefore, the node rankings remain completely unchanged.

\subsection{Network-Level Analysis of the Influence Spreading Model in Random Networks} \label{ism_random}

In this section, we examine the behavior of the influence spreading model on larger randomly generated networks and compare the results to standard centrality metrics. We use scatter plots, Pearson correlation coefficients and Spearman's rank correlation coefficients in our comparisons. The sample Pearson correlation coefficient $r_{xy}$ for a sequence of observation pairs $(x_1, y_1), \dots, (x_n, y_n)$ is defined as
\begin{equation*}
r_{xy} = \frac{\sum_{i=1}^n (x_i - \bar{x})(y_i -\bar{y})}{\sqrt{\left(\sum_{i=1}^n (x_i -\bar{x})^2 \right) \left(\sum_{i=1}^n (y_i -\bar{y})^2 \right)}},
\end{equation*} where $\bar{x} = (1 / n) \sum_{i=1}^n x_i$ and $\bar{y} = (1 / n) \sum_{i=1}^n y_i$ \cite{kaptein2022}. The sample Pearson correlation coefficient is also called the product-moment correlation coefficient.

Spearman's rank correlation coefficient calculates the product moment correlation coefficient between the ranks of pairs of observations \cite{kaptein2022}. If the data contains the same exact values, then ranks for those observations are tied, i.e. they have the same rank. For observations that are tied, Spearman's rank correlation coefficient assigns the average of the untied ranks that would be assigned to these observations if no ties occurred. Formally, Spearman's rank correlation coefficient $r_S$ is defined as \begin{equation*}
r_S = \frac{\sum_{i=1}^n (R^x_i - \bar{R}^x)(R^y_i -\bar{R}^y)}{\sqrt{\left(\sum_{i=1}^n (R^x_i -\bar{R}^x)^2 \right) \left(\sum_{i=1}^n (R^y_i -\bar{R}^y)^2 \right)}},
\end{equation*} where $R^x_i$ and $R^y_i$ are the ranks of observations $x_i$ and $y_i$, respectively, $\bar{R}^x = (1 / n) \sum_{i=1}^n R^x_i$ and $\bar{R}^y = (1 / n) \sum_{i=1}^n R^y_i$. It should be noted that Spearman's rank correlation coefficient does not change if the data is transformed in a monotonic way \cite{kaptein2022}. On the other hand, Pearson correlation coefficient measures the strength of the linear relationship between observations. Therefore, these correlation coefficients can quantify the relationship between observations in different ways.

We have used stochastic network models to generate the random networks used in this section. The stochastic network models used are Erd\H{o}s-R\'enyi random graph model~\cite{erdos1959}, Watts-Strogatz small world model~\cite{watts1998} and Barab\'asi-Albert preferential attachment model~\cite{barabasi1999}. Erd\H{o}s-R\'enyi model comes from the mathematical tradition of random graph theory, whereas Watts-Strogatz and Barab\'asi-Albert models are innovations of network science. These three random network models each capture a different type of network with their characteristic properties and provide an aggregate representation of many individual networks that share these properties. The characteristic properties of the random networks produced by these stochastic network models are very well known, and they have been extensively studied in the literature. Therefore, these three random network models provide varied, yet controlled conditions to assess the influence spreading model and its centrality metrics compared to standard centrality metrics.

We also note that we performed the same or similar experiments on the well-known Les Mis\'erables network~\cite{knuth1993} and on another random network generated by using the extension of the Barab\'asi-Albert model proposed by Holme and Kim~\cite{holme2002}. However, the Les Mis\'erables network data set is quite small with 77 nodes and 254 edges, while similar phenomena appeared in the results as in our chosen three random networks. Therefore, the inclusion of the Les Mis\'erables network was considered redundant. Furthermore, the random network produced by the method proposed by Holme and Kim displayed results similar to Barab\'asi-Albert random network. This is not entirely surprising, since the essential difference between these random network models is that the method of Holme and Kim can create high clustering, whereas the Barab\'asi-Albert model cannot.

The model widely known as the Erd\H{o}s-R\'enyi model (or less widely known as the Erd\H{o}s-R\'enyi-Gilbert model) has two formulations~\cite{erdos1959, gilbert1959}. In the first formulation, the number of nodes and edges is fixed, and then a graph is randomly selected out of all possible graphs. In the second formulation, the number of nodes is fixed, and edges are created between all pairs of nodes independently of all other edges with probability $p$. In this study, we use the second formulation and denote a random Erd\H{o}s-R\'enyi network with $\text{ER}(n, p)$, where $n$ is the number of nodes, and $p$ is the independent edge creation probability. The Erd\H{o}s-R\'enyi model has been thoroughly studied and its properties are well known~\cite{newman2003}. The Erd\H{o}s-R\'enyi model creates networks with low clustering, and its degree distribution has been shown to be the Poisson distribution.

The Watts-Strogatz model uses a regular ring lattice of (fixed number of) $n$ nodes and $k$ neighbors for each node as the starting point, after which edges are rewired with probability $p$~\cite{watts1998}. Thus, with probability $p$, a node is connected to a uniformly randomly selected non-neighbor (duplicate edges are not allowed) and with probability $1 - p$ the edge remains unchanged. With this method of network construction, Watts and Strogatz aimed to develop a mixture of order and randomness to create more realistic networks. The parameter $p$ controls the amount of randomness. The Watts-Strogatz model has the ability to create high clustering and short average path lengths between nodes. Due to the short average path lengths, it is also called the small-world model. The exact degree distribution of the Watts-Strogatz model is derived in~\cite{albert2002}. When $p = 0$, the degree distribution is constant $k$. When $p$ grows, the degree distribution begins to resemble the shape of the degree distribution of the Erd\H{o}s-R\'enyi random graph~\cite{albert2002} i.e.\ the degree distribution becomes approximately Poisson distributed. We denote a network produced by the Watts-Strogatz model by $\text{WS}(n, k, p)$.

The Barab\'asi-Albert model addresses some of the limitations of Erd\H{o}s-R\'enyi model and Watts-Strogatz model~\cite{barabasi1999}. First, the Barab\'asi-Albert model does not assume a fixed number of nodes, and instead, the number of nodes is allowed to grow. Second, Barab\'asi-Albert model does not create edges between nodes with equal probability since a new edge is more likely to be connected to a node that already has a high degree. This is called preferential attachment, where high-degree nodes are more likely to receive more neighbors when the network size grows. The Barab\'asi-Albert model begins from an initial graph and then continues adding nodes using preferential attachment until the desired network size has been reached. We denote the Barab\'asi-Albert network with $\text{BA}(n, m)$, where $n$ is the desired number of nodes, and $m$ is the number of edges to be attached from a new node to existing nodes. The degree distribution of the Barab\'asi-Albert model is scale-free.

Many real-world networks display high clustering and non-Poissonian degree distributions~\cite{newman2003}. Scale-free (i.e. power-law) degree distributions have been proposed to be a common phenomenon for many real networks~\cite{albert2002}. However, the prevalence of scale-free degree distributions in real networks has also been recently disputed in~\cite{broido2019}, where it is actually found that scale-free degree distributions are rare and the log-normal distribution provides an adequate modeling fit instead of a power-law distribution. A short overview on the prevalence of scale-free degree distributions in real networks can be found in~\cite{holme2019}. In light of this differing research on what degree distributions characterize real networks, we argue that testing the behavior of the influence spreading model on random networks that display either Poissonian or scale-free degree distributions is especially justified.

We created three separate random networks with $1000$ nodes using the aforementioned Erd\H{o}s-R\'enyi, Watts-Strogatz and Barab\'asi-Albert models. All random networks were created using the Python package \texttt{NetworkX} (version 2.4)~\cite{SciPyProceedings_11}. The parameters were chosen such that each random network would have roughly the same number of edges. The  random network $\text{ER}(1000, 0.01)$ contains $1000$ nodes and $4910$ edges. The random network $\text{WS}(1000, 10, 0.5)$ has $1000$ nodes and $5000$ edges. The random network $\text{BA}(1000, 5)$ contains $1000$ nodes and $4975$ edges. All three random networks are connected, and were created using seed $1729$.

\begin{figure}[]
\begin{subfigure}{\linewidth}
\includegraphics[scale=0.35]{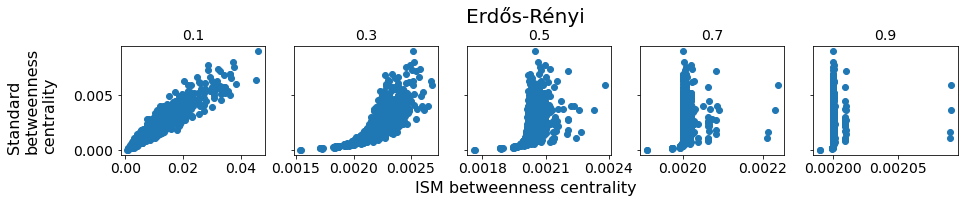} 
\end{subfigure}

\begin{subfigure}{\linewidth}
\includegraphics[scale=0.35]{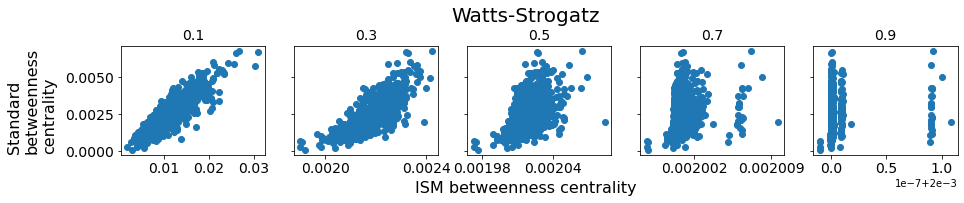} 
\end{subfigure}

\begin{subfigure}{\linewidth}
\includegraphics[scale=0.35]{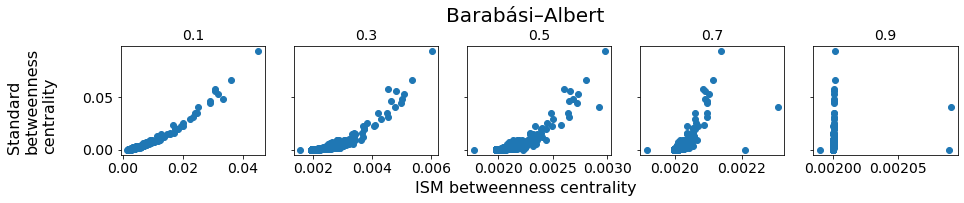} 
\end{subfigure}
\caption{Influence spreading model (with maximum walk length $L_{max} = 20$) betweenness centrality compared to standard betweenness centrality for edge probabilities $0.1, 0.3, 0.5, 0.7$ and $0.9$ on different random networks.} 
\vspace{+6pt}
\label{fig:random_graphs_betweenness}
\end{figure}

In Figure \ref{fig:random_graphs_betweenness}, we have compared ISM betweenness centrality to standard betweenness centrality for three random networks produced by the Erd\H{o}s-R\'enyi, Watts-Strogatz, and Barab\'asi-Albert models. Figure \ref{fig:random_graphs_betweenness} reveals how, for lower edge probabilities, there is a positive linear relationship between the ISM betweenness and the standard betweenness. As the edge probability value increases, this relationship becomes weaker and weaker. Eventually, for the edge probability value $0.9$, we see that multiple levels are formed for the ISM betweenness values. This phenomenon is likely caused by differences in how the network breaks down into components when a node is removed.

In Table \ref{table:corr_ISM_btw_VS_btw} we see that the $\text{BA}(1000, 5)$ network shows quite strong to moderate positive Pearson correlation values for higher edge probabilities, whereas for the $\text{ER}(1000, 0.01)$ and $\text{WS}(1000, 10, 0.5)$ networks the correlation has significantly weakened. For the $\text{BA}(1000, 5)$ network and for the highest edge probabilities, Pearson correlation becomes weak as well.
\begin{table}[b]
\vspace{-6pt}
\caption{Pearson correlation coefficient between ISM betweenness centrality and standard betweenness centrality.}
\begin{tabular}{|p{3cm}|p{0.9cm}|p{0.9cm}|p{0.9cm}|p{0.9cm}|p{0.9cm}|p{0.9cm}|p{0.9cm}|p{0.9cm}|p{0.9cm}| }
 \hline
 \multicolumn{1}{|c|}{} &
 \multicolumn{9}{|c|}{Edge probability} \\
 \cline{2-10}
 Network & 0.1 & 0.2 & 0.3 & 0.4 & 0.5 & 0.6 & 0.7 & 0.8 & 0.9 \\
 \hline
 $\text{ER}(1000, 0.01)$ & 0.929 & 0.939 & 0.803 & 0.630 & 0.451 & 0.296 & 0.182 & 0.107 & 0.061 \\
 $\text{WS}(1000, 10, 0.5)$ & 0.887 & 0.891 & 0.787 & 0.665 & 0.527 & 0.380 & 0.250 & 0.157 & 0.105 \\
 $\text{BA}(1000, 5)$ & 0.964 & 0.883 & 0.857 & 0.847 & 0.839 & 0.819 & 0.727 & 0.424 & 0.171 \\
 \hline
\end{tabular}
\label{table:corr_ISM_btw_VS_btw}
\end{table}

In Table \ref{table:Spearman_corr_ISM_btw_VS_btw}, we see that for the $\text{BA}(1000, 5)$ network Spearman's rank correlation remains quite high for all edge probabilities. However, for the $\text{ER}(1000, 0.01)$ and $\text{WS}(1000, 10, 0.5)$ networks Spearman's rank correlation becomes weaker as edge probability increases. A notable difference to Pearson correlation is that for the highest edge probabilities there is more correlation between the ranks than there is between the values.
\begin{table}[]
\caption{Spearman's rank correlation coefficient between ISM betweenness centrality and standard betweenness centrality.}
\begin{tabular}{|p{3cm}|p{0.9cm}|p{0.9cm}|p{0.9cm}|p{0.9cm}|p{0.9cm}|p{0.9cm}|p{0.9cm}|p{0.9cm}|p{0.9cm}| }
 \hline
 \multicolumn{1}{|c|}{} &
 \multicolumn{9}{|c|}{Edge probability} \\
 \cline{2-10}
 Network & 0.1 & 0.2 & 0.3 & 0.4 & 0.5 & 0.6 & 0.7 & 0.8 & 0.9 \\
 \hline
 $\text{ER}(1000, 0.01)$ & 0.936 & 0.979 & 0.877 & 0.721 & 0.590 & 0.506 & 0.456 & 0.437 & 0.429 \\
 $\text{WS}(1000, 10, 0.5)$ & 0.871 & 0.887 & 0.796 & 0.678 & 0.553 & 0.446 & 0.363 & 0.313 & 0.269 \\
 $\text{BA}(1000, 5)$ & 0.915 & 0.923 & 0.911 & 0.896 & 0.878 & 0.860 & 0.844 &  0.834 & 0.831 \\
 \hline
\end{tabular}
\label{table:Spearman_corr_ISM_btw_VS_btw}
\end{table}

Interestingly, the extension of the Barab\'asi-Albert model proposed by Holme and Kim produced a random network that maintained a strong Pearson correlation between ISM betweenness and standard betweenness even for the highest edge probability value $0.9$. Stronger positive Pearson correlation between ISM betweenness and standard betweenness for the edge probabilities $0.7, 0.8$, and $0.9$ was the only notable difference of the Holme-Kim network compared to the $\text{BA}(1000, 5)$ network in our experiments.

\begin{figure}[t]
\begin{subfigure}{\linewidth}
\includegraphics[scale=0.35]{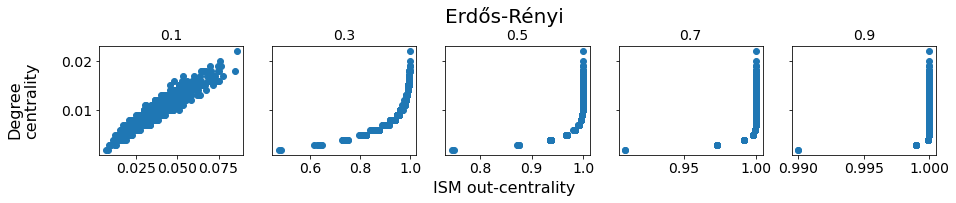} 
\end{subfigure}

\begin{subfigure}{\linewidth}
\includegraphics[scale=0.35]{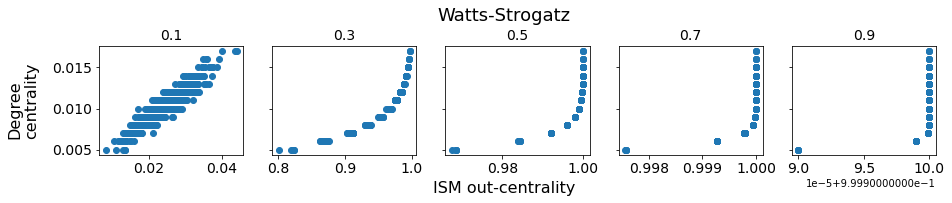} 
\end{subfigure}

\begin{subfigure}{\linewidth} 
\includegraphics[scale=0.35]{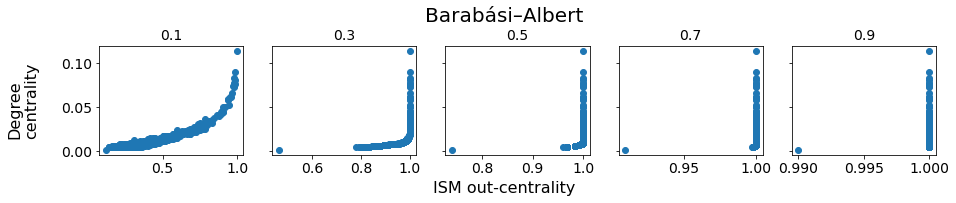} 
\end{subfigure}
\caption{Influence spreading model (with maximum walk length $L_{max} = 20$) out-centrality compared to degree centrality for edge probabilities $0.1, 0.3, 0.5, 0.7$, and $0.9$ on random networks.} \label{fig:random_graphs_out-centrality_degree}
\end{figure}

In Figure \ref{fig:random_graphs_out-centrality_degree}, out-centrality is compared to degree centrality. The Barab\'asi-Albert model creates a heavy-tailed degree distribution (scale-free distributions are heavy-tailed), resulting in hubs. A hub is a node with a great number of neighbors compared to the average degree. In this particular 1000-node Barab\'asi-Albert network, there is a node with more than 100 neighbors. This and other hubs cause a non-linear relationship between out-centrality and degree centrality values even for the edge probability value $0.1$.

Another notable phenomenon in Figure \ref{fig:random_graphs_out-centrality_degree} is how the scale of out-centrality values changes for different networks as the probability of the edges increases. Both networks $\text{ER}(1000, 0.01)$ and $\text{WS}(1000, 10, 0.5)$ display very small out-centrality values for the edge probability value $0.1$. However, for higher edge probabilities, some nodes attain out-centrality values that are close to one (maximum value for out-centrality), and as the edge probability increases, the interval of out-centrality values begins to narrow as minimum values approach one. For the $\text{BA}(1000, 5)$ network, the pattern is similar with one notable exception. For the edge probability value $0.1$, the hub nodes of this network already achieve values that are close to the maximum out-centrality value.

In Table \ref{table:corr_out_degree}, the Pearson correlation coefficient between out-centrality and degree centrality is shown for all edge probabilities. For the edge probability value $0.1$, the $\text{ER}(1000, 0.01)$ and $\text{WS}(1000, 10, 0.5)$ networks display a quite strong positive Pearson correlation, which then becomes weaker as the edge probability increases. For the $\text{BA}(1000, 5)$ network, there is also a positive Pearson correlation. However, the non-linear relationship makes the Pearson correlation weaker. In Table \ref{table:Spearman_corr_out_degree}, we see that Spearman's rank correlation remains high for all random networks as the edge probability increases. We can infer that as edge probability increases for all edges simultaneously, the relative rankings between nodes do not change much.
\begin{table}[]
\caption{Pearson correlation coefficient between out-centrality and degree centrality.}
\begin{tabular}{|p{3cm}|p{0.9cm}|p{0.9cm}|p{0.9cm}|p{0.9cm}|p{0.9cm}|p{0.9cm}|p{0.9cm}|p{0.9cm}|p{0.9cm}| }
 \hline
 \multicolumn{1}{|c|}{} &
 \multicolumn{9}{|c|}{Edge probability} \\
 \cline{2-10}
 Network & 0.1 & 0.2 & 0.3 & 0.4 & 0.5 & 0.6 & 0.7 & 0.8 & 0.9 \\
 \hline
 $\text{ER}(1000, 0.01)$ & 0.950 & 0.928 & 0.821 & 0.690 & 0.552 & 0.422 & 0.310 & 0.222 & 0.157 \\
 $\text{WS}(1000, 10, 0.5)$ & 0.935 & 0.962 & 0.905 & 0.817 & 0.702 & 0.567 & 0.431 & 0.315 & 0.232 \\
 $\text{BA}(1000, 5)$ & 0.865 & 0.681 & 0.573 & 0.491 & 0.406 & 0.289 & 0.154 & 0.063 & 0.030 \\
 \hline
\end{tabular}
\vspace{+6pt}
\label{table:corr_out_degree}
\end{table}
\begin{table}[]
\vspace{-12pt}
\caption{Spearman's rank correlation coefficient between out-centrality and degree centrality.}
\begin{tabular}{|p{3cm}|p{0.9cm}|p{0.9cm}|p{0.9cm}|p{0.9cm}|p{0.9cm}|p{0.9cm}|p{0.9cm}|p{0.9cm}|p{0.9cm}| }
 \hline
 \multicolumn{1}{|c|}{} &
 \multicolumn{9}{|c|}{Edge probability} \\
 \cline{2-10}
 Network & 0.1 & 0.2 & 0.3 & 0.4 & 0.5 & 0.6 & 0.7 & 0.8 & 0.9 \\
 \hline
 $\text{ER}(1000, 0.01)$ & 0.947 & 0.992 & 0.995 & 0.995 & 0.995 & 0.995 & 0.995 & 0.995 & 0.997 \\
 $\text{WS}(1000, 10, 0.5)$ & 0.928 & 0.987 & 0.987 & 0.987 & 0.987 & 0.987 & 0.987 & 0.987 & 0.999 \\
 $\text{BA}(1000, 5)$ & 0.848 & 0.976 & 0.985 & 0.985 & 0.985 & 0.985 & 0.985 & 0.985 & 0.984 \\
 \hline
\end{tabular}
\label{table:Spearman_corr_out_degree}
\end{table}

Figure \ref{fig:random_graphs_out-centrality_degree} and Table \ref{table:corr_out_degree} show that for low edge probabilities, the degree is a good predictor for out-centrality values when the degree distribution is Poissonian. When the degree distribution is scale-free, a linear relationship does not fully capture the dynamics between out-centrality and degree centrality. For higher edge probabilities, all the random networks begin to display a non-linear relationship between out-centrality and degree centrality. $\text{ER}(1000, 0.01)$ and $\text{WS}(1000, 10, 0.5)$ networks maintain a stronger positive correlation with degree centrality for much higher edge probabilities than $\text{BA}(1000, 5)$ network. The difference again comes from the light-tailed Poisson degree distribution that is shared by the $\text{ER}(1000, 0.01)$ and $\text{WS}(1000, 10, 0.5)$ networks, whereas $\text{BA}(1000, 5)$ network has a heavy-tailed degree distribution.

\begin{figure}[t]
\begin{subfigure}{\linewidth}
\includegraphics[scale=0.35]{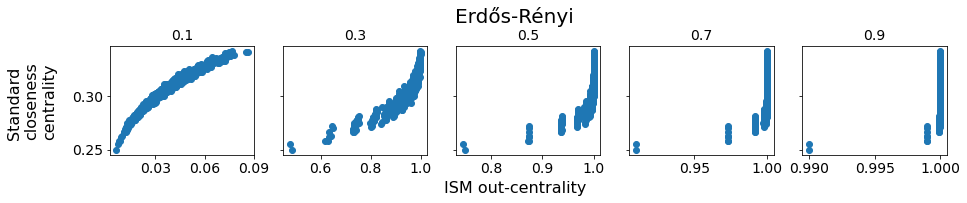} 
\end{subfigure}

\begin{subfigure}{\linewidth}
\includegraphics[scale=0.35]{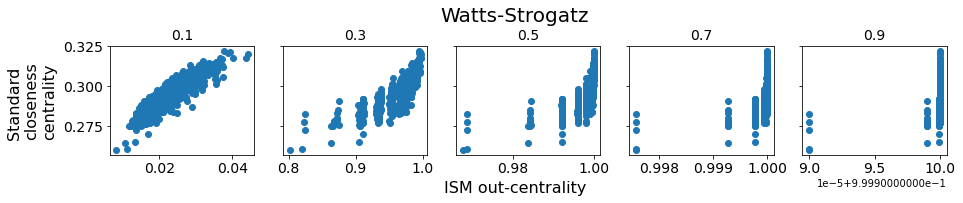} 
\end{subfigure}

\begin{subfigure}{\linewidth} 
\includegraphics[scale=0.35]{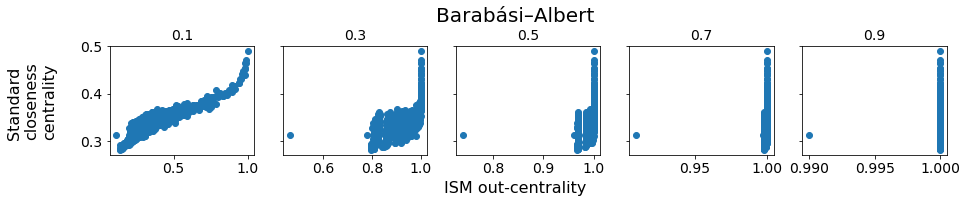} 
\end{subfigure}
\caption{Influence spreading model (with maximum walk length $L_{max} = 20$) out-centrality compared to standard closeness centrality for edge probabilities $0.1, 0.3, 0.5, 0.7$, and $0.9$ on random networks.} \label{fig:random_graphs_out-centrality_closeness}
\end{figure}

In Figure \ref{fig:random_graphs_out-centrality_closeness},  the out-centrality is compared to the standard closeness centrality for $\text{ER}(1000, 0.01)$, $\text{WS}(1000, 10, 0.5)$, and $\text{BA}(1000, 5)$ random networks. Again, it is important to notice the changing scale of out-centrality values as the edge probability increases. Out-centrality and standard closeness centrality display a similar relationship as seen between out-centrality and degree centrality.

In Table \ref{table:corr_out_closeness}, we see a positive Pearson correlation for low edge probabilities, which becomes weaker as the edge probability increases, and eventually the correlation is close to zero. As edge probability increases, $\text{ER}(1000, 0.01)$ and $\text{WS}(1000, 10, 0.5)$ networks maintain a stronger positive Pearson correlation than $\text{BA}(1000, 5)$ network. We see that for low edge probabilities, closeness centrality is a good predictor for out-centrality values. We can see that for low edge probabilities, the close and immediate neighborhood of a node determines its ability to influence other nodes (out-centrality), since the low influence probability decays significantly over longer walk distances.

\begin{table}[]
\caption{Pearson correlation coefficient between out-centrality and standard closeness centrality.}
\begin{tabular}{|p{3cm}|p{0.9cm}|p{0.9cm}|p{0.9cm}|p{0.9cm}|p{0.9cm}|p{0.9cm}|p{0.9cm}|p{0.9cm}|p{0.9cm}| }
 \hline
 \multicolumn{1}{|c|}{} &
 \multicolumn{9}{|c|}{Edge probability} \\
 \cline{2-10}
 Network & 0.1 & 0.2 & 0.3 & 0.4 & 0.5 & 0.6 & 0.7 & 0.8 & 0.9 \\
 \hline
 $\text{ER}(1000, 0.01)$ & 0.974 & 0.959 & 0.871 & 0.757 & 0.630 & 0.502 & 0.385 & 0.288 & 0.212 \\
 $\text{WS}(1000, 10, 0.5)$ & 0.893 & 0.873 & 0.817 & 0.745 & 0.652 & 0.541 & 0.425 & 0.323 & 0.248 \\
 $\text{BA}(1000, 5)$ & 0.917 & 0.762 & 0.649 & 0.561 & 0.467 & 0.335 & 0.179 & 0.073 & 0.033 \\
 \hline
\end{tabular}
\label{table:corr_out_closeness}
\vspace{+12pt}
\end{table}

In Table \ref{table:Spearman_corr_out_closeness}, we can see that Spearman's rank correlation remains relatively high compared to Pearson correlation for all random networks as the edge probability increases. Interestingly, when edge probabilities are high, the $\text{BA}(1000, 5)$ network has lower rank correlation than the other two networks.

\begin{figure}[]
\begin{subfigure}{\linewidth}
\includegraphics[scale=0.34]{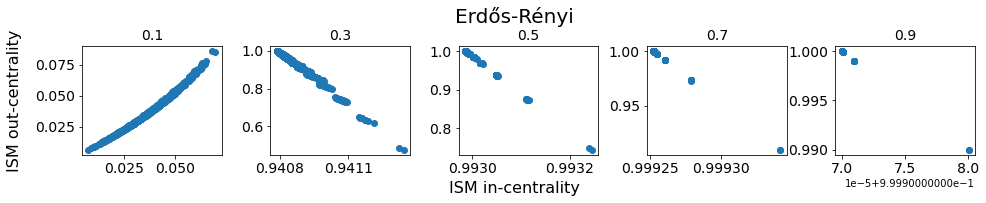} 
\end{subfigure}

\begin{subfigure}{\linewidth}
\includegraphics[scale=0.34]{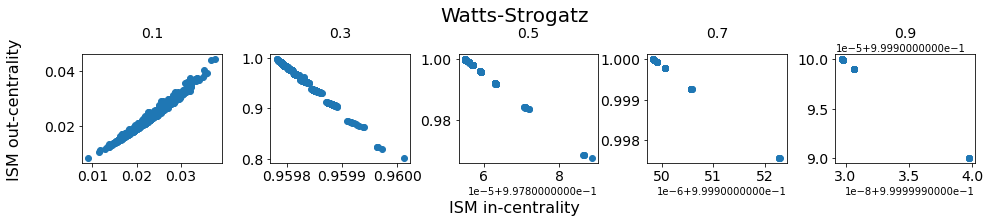} 
\end{subfigure}

\begin{subfigure}{\linewidth} 
\includegraphics[scale=0.34]{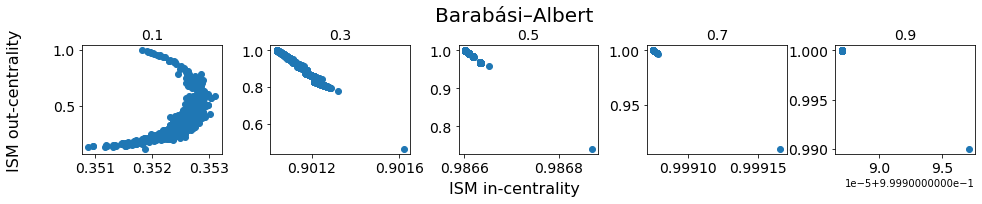} 
\end{subfigure}
\caption{Influence spreading model (with maximum path length $L = 20$) in-centrality compared to out-centrality for edge probabilities $0.1, 0.3, 0.5, 0.7$, and $0.9$ on random networks.} 
\vspace{-2pt}
\label{fig:random_graphs_out-centrality_in-centrality}
\end{figure}

In Figure \ref{fig:random_graphs_out-centrality_in-centrality}, the relationship between in-centrality and out-centrality is studied for the $\text{ER}(1000, 0.01)$, $\text{WS}(1000, 10, 0.5)$, and $\text{BA}(1000, 5)$ networks. It is important to note that both the x-axis and y-axis scale changes in every individual plot. We see the same behavior for out-centrality and in-centrality values that was seen for the Krackhardt kite network in Section \ref{krackhardt}. Out-centrality values have much larger dispersion, whereas there is a small variation between in-centrality values.

\begin{table}[b]
\vspace{-6pt}
\caption{Spearman's rank correlation coefficient between out-centrality and closeness centrality.}
\begin{tabular}{|p{3cm}|p{0.9cm}|p{0.9cm}|p{0.9cm}|p{0.9cm}|p{0.9cm}|p{0.9cm}|p{0.9cm}|p{0.9cm}|p{0.9cm}| }
 \hline
 \multicolumn{1}{|c|}{} &
 \multicolumn{9}{|c|}{Edge probability} \\
 \cline{2-10}
 Network & 0.1 & 0.2 & 0.3 & 0.4 & 0.5 & 0.6 & 0.7 & 0.8 & 0.9 \\
 \hline
 $\text{ER}(1000, 0.01)$ & 0.991 & 0.976 & 0.968 & 0.965 & 0.963 & 0.962 & 0.961 & 0.960 & 0.952 \\
 $\text{WS}(1000, 10, 0.5)$ & 0.892 & 0.881 & 0.877 & 0.874 & 0.872 & 0.869 & 0.867 & 0.866 & 0.849 \\
 $\text{BA}(1000, 5)$ & 0.901 & 0.735 & 0.701 & 0.693 & 0.687 & 0.684 & 0.682 & 0.682 & 0.673 \\
 \hline
\end{tabular}
\label{table:Spearman_corr_out_closeness}
\end{table}

In Figure \ref{fig:random_graphs_out-centrality_in-centrality} and in Table \ref{table:corr_in_out}, both $\text{ER}(1000, 0.01)$ and $\text{WS}(1000, 10, 0.5)$ networks display an initially strong positive Pearson correlation, which eventually reverses into a strong negative Pearson correlation as the edge probability increases. However, for the $\text{BA}(1000, 5)$ network, the reversal has already begun for the edge probability value $0.1$ since part of the scatter plot shows a positive correlation, while the other part shows a negative correlation. As the edge probability increases, the correlation between in-centrality and out-centrality becomes strongly negative. We are again essentially seeing the inverting ranking behavior of out-centrality and in-centrality as the edge probability increases. In Table \ref{table:Spearman_corr_in_out}, this inverted ranking can be verified by looking at the change in rank correlation as edge probability increases.

In our experiments we have discovered that network-wide inverting ranking behavior between out-centrality and in- centrality does not occur when some nodes are completely disconnected from the rest of the network. For such a disconnected (zero degree) node, its in-centrality and out-centrality equal zero for all edge probabilities. The rest of the network may exhibit a reversing behavior between in-centrality and out-centrality, although nodes without neighbors disrupt the change in correlation from positive to negative as their values remain static at zero.

\begin{table}[]
\caption{Pearson correlation coefficient between in-centrality and out-centrality.}
\begin{tabular}{|p{3cm}|p{0.9cm}|p{0.9cm}|p{0.9cm}|p{0.9cm}|p{0.9cm}|p{0.9cm}|p{0.9cm}|p{0.9cm}|p{0.9cm}| }
 \hline
 \multicolumn{1}{|c|}{} &
 \multicolumn{9}{|c|}{Edge probability} \\
 \cline{2-10}
 Network & 0.1 & 0.2 & 0.3 & 0.4 & 0.5 & 0.6 & 0.7 & 0.8 & 0.9 \\
 \hline
 $\text{ER}(1000, 0.01)$ & 0.995 & -0.984 & -0.998 & -0.999 & -1.000 & -1.000 & -1.000 & -1.000 & -1.000 \\
 $\text{WS}(1000, 10, 0.5)$ & 0.992 & -0.987 & -0.998  & -1.000 & -1.000 & -1.000 & -1.000 & -1.000 & -1.000 \\
 $\text{BA}(1000, 5)$ & 0.532 & -0.972 & -0.995 & -0.999 & -1.000 & -1.000 & -1.000 & -1.000 & -1.000 \\
 \hline
\end{tabular}
\label{table:corr_in_out}
\end{table}
\vspace{-6pt}
\begin{table}[H]
\caption{Spearman's rank correlation coefficient between in-centrality and out-centrality.}
\begin{tabular}{|p{3cm}|p{0.9cm}|p{0.9cm}|p{0.9cm}|p{0.9cm}|p{0.9cm}|p{0.9cm}|p{0.9cm}|p{0.9cm}|p{0.9cm}| }
 \hline
 \multicolumn{1}{|c|}{} &
 \multicolumn{9}{|c|}{Edge probability} \\
 \cline{2-10}
 Network & 0.1 & 0.2 & 0.3 & 0.4 & 0.5 & 0.6 & 0.7 & 0.8 & 0.9 \\
 \hline
 $\text{ER}(1000, 0.01)$ & 1.000 & -0.989 & -0.999 & -1.000 & -1.000 & -1.000 & -1.000 & -0.999 & -0.966 \\
 $\text{WS}(1000, 10, 0.5)$ & 0.994 & -0.992 & -1.000 & -1.000 & -1.000 & -1.000 & -1.000 & -0.992 & -0.976 \\
 $\text{BA}(1000, 5)$ & 0.828 & -0.958 & -0.999 & -1.000 & -1.000 & -1.000 & -1.000 & -0.999 & -0.990 \\
 \hline
\end{tabular}
\label{table:Spearman_corr_in_out}
\end{table}

\section{Discussion} \label{discussion}

The influence spreading model provides a flexible framework for studying a wide range of complex networks, as it supports directed and undirected edges as well as arbitrary edge probabilities. In this study, we focus on undirected networks with equal edge probabilities, which provides a controlled setting to isolate the effects of edge probability variation and understand the behavior of the model as this parameter changes.

All experiments in Sections~\ref{krackhardt} and~\ref{ism_random} were conducted using a maximum walk length of $L_{max}=20$. Additional experiments with shorter and longer walk lengths show that allowing longer walks causes the reversing behavior in in-centrality rankings to occur earlier, whereas restricting walk length delays this effect. In general, increasing $L_{max}$ raises the probability of influence by permitting a greater number of walks. The impact of varying the maximum allowed walk length has been discussed in detail in~\cite{kuikka2022efficiency}, where the term path was used with the same meaning as walk in this article.

This study considers only the complex contagion version of the influence spreading model. For random networks generated using the Erd\H{o}s–R\'enyi, Watts– Strogatz, and Barab\'asi–Albert models, in-centrality and out-centrality are positively correlated at low edge probabilities. As the edge probability increases, this relationship reverses and becomes negative. This phenomenon cannot occur under simple contagion, where cycles are absent and the number of paths between node pairs is symmetric. In complex contagion, cycles act as amplifiers, which in social networks can be interpreted as echo chambers.

When networks are undirected and edge probabilities are identical, the in-centrality and out-centrality are determined solely by network topology. Peripheral nodes tend to have lower out-centrality, while more central nodes exhibit greater influence, a pattern that persists as edge probability increases. For low edge probabilities, the peripheral nodes are also the least influenced, whereas the central nodes are the most influenced. As edge probability increases, this pattern reverses, with peripheral nodes becoming the most influenced and central nodes the least influenced. This behavior is consistent with the intuitive properties of real social networks, where the central nodes are typically more influential and more exposed to influence.

The proposed ISM betweenness centrality reveals structural properties that standard betweenness centrality cannot reveal, such as the importance of alternative routes and critical bridging nodes. However, this increased expressiveness comes at a high computational cost, as the influence spreading algorithm must be recomputed for each node removal. Thus, while ISM betweenness captures more of the underlying network topology, it does so at the expense of computational efficiency.

Our study reveals similar behavior between the centrality measures proposed by the influence spreading model and the standard centrality measures. However, the influence spreading model also allows for in-built probability interpretation for the values of centrality measures. In some application areas, this natural probability interpretation can be more valuable than the values provided by other standard centrality measures. Our study also highlights that the values and rankings calculated from different centrality measures would reveal different information.

\section{Conclusion} \label{conclusions}

Many standard centrality measures, including closeness and betweenness centrality, rely on shortest paths. In contrast, the influence spreading model defines centrality metrics based on unrestricted paths, walks, and cycles. For low edge probabilities, the resulting metrics exhibit similarities to standard centrality measures, while also revealing structural information that shortest-path-based metrics cannot capture. The main contribution of this study is a detailed analysis of ISM betweenness centrality, out-centrality, and in-centrality compared to standard centrality measures.

We showed that ISM betweenness centrality highlights the importance of alternative routes while remaining comparable to standard betweenness centrality. In addition, out-centrality identifies influential central nodes, whereas in-centrality highlights peripheral nodes as the most influenced when network edges are undirected and edge probabilities are uniform. These results support the validity of the proposed centrality measures and suggest that in more realistic networks with directed edges and heterogeneous edge probabilities, the distinctions between node roles will become even more pronounced.

\renewcommand{\thesection}{A}
\appendix
\section{Appendix} \label{appendix}

The influence spreading model uses probabilities as edge weights. However, standard betweenness centrality and closeness centrality require a distance to be defined between nodes. In real-world networks, there is some sort of relationship between the influence spreading probability and the distance between nodes. For example, during the COVID-19 pandemic, social distancing was encouraged in order to reduce the probability of infection. Therefore, there is a need to map edge probabilities to edge distances in a systematic way. This is especially true if the edge probabilities are not the same for all edges.

One possible interpretation of a shortest path in a graph that has probabilities as edge weights is to interpret the path with the highest probability as if it were the shortest path. If we denote $p_1, p_2, \dots, p_m$ as the edge probabilities in a graph $G = (V, E)$, where $V$ is the set of nodes, $E$ is the set of edges between nodes, and $m = |E|$ is the number of edges, then our task is to find a path between source node $s$ and target node $t$ that maximizes the product $p_{(s, a)} \cdots p_{(b, t)}$, where $(s, a), \dots, (b, t) \in E$ are the edges and $s, a, \ldots, b, t \in V$ are the nodes on the path that maximizes the probability of influence between nodes $s$ and $t$.

Maximizing the product $p_{(s, a)} \cdots p_{(b, t)}$ is equivalent to maximizing the sum $\log p_{(s, a)} + \ldots + \log p_{(b, t)}$, which is equivalent to minimizing $-\log p_{(s, a)} - \ldots - \log p_{(b, t)}$, which is again equivalent to minimizing $\log (1 /p_{(s, a)}) + \ldots + \log (1 / p_{(b, t)}).$ In the previous formulation, we assumed that the shortest path contains at least one intermediary node. If that is not the case, then naturally the shortest path between $s$ and $t$ is the edge $(s, t)$. 

When $0 < p_i \leq 1$ for all $1 \leq i \leq m$, then $\log (1 / p_i) \geq 0$ for all $1 \leq i \leq m$. Now, $\log (1 / p_i)$, for all $1 \leq i \leq m$, can be used as a nonnegative distance between nodes, because as probability $p$ increases the distance $\log (1 / p)$ decreases, i.e.\ $\log (1 / p)$ is a strictly decreasing function of $p$. Furthermore, we can now use Dijkstra's algorithm for finding shortest paths in the network, since edge weights are non-negative~\cite{thomas2009}. In this case, Dijkstra's algorithm picks the edges with the lowest distance $\log (1 / p_i)$. Therefore, with this transformation the edge $i$ with the lowest value $\log (1 / p_i)$ is chosen i.e.\ the highest probability edge is chosen for the shortest path. In the special case, where $p_i = 1$ for all $1 \leq i \leq m$, then the problem reduces to finding the path with the fewest edges.

%
%

\printbibliography

\end{document}